\documentclass[prl,twocolumn,graphicx,amssymb,floatfix]{revtex4-1}
\usepackage{graphicx}
\usepackage{color}
\usepackage{soul}
\usepackage{float}
\begin{document}

\title{A Comment on ``Paradox of photons disconnected trajectories being located by means of ``weak measurements'' in the nested Mach-Zehnder interferometer''}
\author{ Lev Vaidman}
\affiliation{  Raymond and Beverly Sackler School of Physics and Astronomy,
 Tel-Aviv University, Tel-Aviv 69978, Israel
}

\noindent

\begin{abstract}
The criticism of the experiment showing discontinuous traces of photons passing through a nested Mach-Zehnder interferometer is shown to be unfounded.
\end{abstract}
\maketitle
In a recent Letter \cite{Nik} Nikolaev reviewed an experiment by Danan et al. \cite{Danan} which demonstrated the ``past of a photon'' defined as a weak trace left  by a pre- and post-selected photon. The experiment was already widely discussed in the literature \cite{LiCom,pastReply,pastsec,Saldaha,china_experiment,improved,PF,comment_potocek,Salih,comment_salih,Bart,ComBart,RepComBart,Bula,Hashimi,reply-hashimi,Wu,Grif,reply-Grif,Sven,Repl-Sven}, but Nikolaev claimed to add new considerations which show  ``in a maximally clear way'' that it can be fully explained by the traditional wave theory of light or by quantum theory assuming continuous trajectories of photons. He argued that the surprising discontinuous trace in the inner interferometer appeared due to the improper way of its observation and that the missing parts of the trace  can be revealed by modifying the original setup.

In this Comment I will argue that it is the modification of Nikolaev which is improper and that the original experiment faithfully demonstrates the  discontinuous local trace of the pre- and post-selected photons.

There is no controversy about the possibility of explaining the results of Danan et al.'s experiment using classical equations of electromagnetic waves. This was explained in the original paper  itself, mostly in Supplementary material.  Moreover, an explanation in the framework of  standard quantum theory was given in the original paper. This explanation, however, required an analysis of the entanglement with the measuring device and was not based on the photon's continuous trajectories. Standard quantum theory does not have the concept of photon trajectory. Vaidman \cite{past} defined the past of a quantum particle as places where it leaves a weak trace and Danan et al.'s experiment demonstrated it.

A direct observation of a weak trace of a pre- and post-selected particle requires an external measuring device and a mechanism which takes the record of the device into account only on the condition of a successful post-selection  measurement \cite{improved}.  Such an experiment, however, is very demanding - see recent results by Steinberg \cite{Steinberg} - and in  most weak measurements the measuring device is another degree of freedom of the particle itself. This avoids the need for the mechanism arranging coincidence counting.  In Danan et al.'s experiment the record of local weak coupling, the small shift in the direction of propagation due to the rotation of the internal mirrors,  takes place inside the interferometer, but the reading is performed outside the interferometer at the quad-cell detector. This was made possible  by the careful design of the experiment which ensured that the shift of the direction in the region of the local coupling is translated to the shift of the output beam in the same way for all possible paths of the beam toward the detector. Without this property the signal at the detector does not present a faithful indication of the trace inside the interferometer.

Nikolaev claims  that the reason Danan et al.'s  experiment showed no signal corresponding to the presence of the photon in $E$ was its improper design. He notices that modulation of the polarization in $E$ would lead to an observable signal if a birefringent plate and a polarizer are added in one of the arms of the inner interferometer. But it is Nikolaev's  modification that makes the design improper  since it causes different transformations of the polarization record at $E$ into a signal at the detector, depending on the path the beam takes.

In fact, similar behavior has been noticed before \cite{Jordan}. Introducing a Dove prism in one of the arms of the original interferometer \cite{Danan} leads to sensitivity to the angle modulation of the mirror at $E$. This is because the Dove prism flips the response at the  detector to a change of the beam direction at $E$, spoiling the necessary property of identical responses for all possible paths.

Direct measurement of the weak trace using an external measuring device would show the results observed by Danan et al. The photon leaves a trace in a continuous path $C$ but also, separately, inside the inner interferometer which includes mirrors $A$ and $B$.

It is true that Danan et al.'s experiment can be fully explained using interference of classical electromagnetic waves. It can also be explained by the photon's wave function being entangled with the measuring device (which is the transversal momentum of the photon  in Danan et al.'s experiment). But if one wants to ask where a pre- and post-selected particle has been, defining it as the places where it left a local trace, there is no better proposal than the simple criterion of the overlap of the forward and backward evolving quantum states of the two-state vector formalism \cite{AV90}.

This work has been supported in part by the Israel Science Foundation Grant No. 1311/14,
the German-Israeli Foundation for Scientific Research and Development Grant No. I-1275-303.14.

\end{document}